\documentclass[prl,twocolumn,showpacs,amsmath,amssymb]{revtex4}

\usepackage{graphicx}
\usepackage{bm}
\usepackage{epsfig,psfrag,graphics}

\begin{document}


\title{Dynamical Heterogeneity close to the Jamming Transition \\in a Sheared Granular Material}

\author{O. Dauchot}
\affiliation{SPEC, CEA-Saclay, 91 191 Gif-sur-Yvette, France}
\author{G. Biroli}
\affiliation{SPhT, CEA-Saclay, 91 191 Gif-sur-Yvette, France}
\author{G. Marty}
\affiliation{SPEC, CEA-Saclay, 91 191 Gif-sur-Yvette, France}

\date{\today}

\begin{abstract}
The dynamics of a bi-dimensional dense granular packing under cyclic shear is experimentally investigated close to the jamming transition. Measurement of multi-point correlation functions are produced. The self-intermediate scattering function, displaying slower than exponential relaxation, suggests dynamic heterogeneity. Further analysis of four point correlation functions reveal that the grain relaxations are strongly correlated and spatially heterogeneous, especially at the time scale of the collective rearrangements. Finally, a dynamical correlation length is extracted from spatio-temporal pattern of mobility. Our experimental results open the way to a systematic study of dynamic correlation functions in granular materials.  
\end{abstract}

\maketitle
The dynamical behavior of granular media close to the 'jamming transition' is very similar to the one of liquids close to the glass transition~\cite{LiuNagel}. Indeed granular media close to jamming display a similar dramatic slowing down of the dynamics~\cite{D'Anna,Compaction} as well as other glassy features such as aging and memory effect \cite{aging}. 
Recently, a ``microscopic'' confirmation of the above similarity has been obtained analyzing directly the grain dynamics under cyclic shear during compaction~\cite{Pouliquen} or at constant density~\cite{Marty}. The typical trajectories of grains display the so-called cage effect and are remarkably similar to the ones observed in experiments on colloidal suspension~\cite{Weeks} and in molecular dynamics simulations of glass-formers~\cite{KA}.
As for glass-formers, and contrary to standard critical slowing down, this
slow glassy dynamics does not seem related to a growing {\it static} local
order. For glass-formers it has been shown numerically~\cite{Andersen,Harrowell,Glotzer,Berthier} and experimentally~\cite{Ediger} that instead the {\it dynamics} becomes strongly heterogeneous and {\it dynamic correlations} build up when approaching  the glass transition. The existence of a growing dynamic correlation length is very important to reveal some kind of criticality
associated to the glass transition \cite{Ball}.\\
Here we unveil that also granular materials are strongly dynamically
correlated close to the jamming transition. First, we shall focus on two point functions, in particular the self-intermediate scattering function, whose slower than exponential relaxation suggests dynamic heterogeneity.
Then, following recent theoretical suggestions~\cite{Lefevre,Arenzon}, we shall turn to four point correlation functions. They have been introduced for glass-formers to measure properly dynamic correlations \cite{FranzParisi} and indeed reveal that the dynamics is strongly correlated and heterogeneous. Finally, we shall focus on spatio-temporal pattern of mobility, out of which we extract a direct measurement of a dynamical lengthscale.
Our experimental results, to our knowledge the first direct measurement of
four point {\it spatio-temporal} correlation functions \cite{fourpoint}, open the way to a systematic study of dynamic correlation functions in granular material as a way towards a better understanding of glassy and "jammy" materials in general.\\ 
The experimental setup, a more robust and better designed version of the one presented in \cite{Marty} is as follows : a bi-dimensional, bi-disperse granular material, composed of about 8.000 metallic cylinders of diameter 5 and 6 mm in equal proportions, is sheared quasi-statically in an horizontal deformable parallelogram.
The shear is periodic, with an amplitude $\theta_{max} = \pm 5^{\circ}$. The volume fraction $(\Phi = 0.84)$ is maintained constant by imposing the height of the parallelogram. We follow $2818$ grains located in the center of the device to avoid boundary effects with a High Resolution Digital Camera which takes a picture each time
the system is back to its initial position $\theta = 0$. The unit of time is
one cycle, a whole experiment lasting 10.000 cycles.  The unit of length
is chosen to be the mean particle diameter $d$. These conditions are very
similar to \cite{Marty} and by repeating the same analysis we find a cage
radius of $0.2$ and a cage lifetime of $300$. As discussed in \cite{Marty}, the diffusion is isotropic, at least far from the borders of the experimental cell.\\
Let us first focus on the self intermediate scattering function which measures the dynamics of single particles:
\begin{equation} \label{self}
F_s=\langle \hat F_s(k,t) \rangle=\frac 1 N \sum_j \langle \exp\left[-i { k}(r_j(t)-r_j(0)) \right] \rangle,
\end{equation}
where $r_j(t)$ is the position of the jth particles at time $t$. $\hat F_s(k,t)$ denotes the non averaged instantaneous observable. $\langle \cdot \rangle$ means (here and in the following) a time average over $300$ steps of $10$ cycle each computed after a few thousand cycles, when the systems has reached a steady state (at least on the timescale of the experiment). The sum in (\ref{self}) is over all tracked particles.\\        
\begin{figure}
\includegraphics[width=0.9\columnwidth]{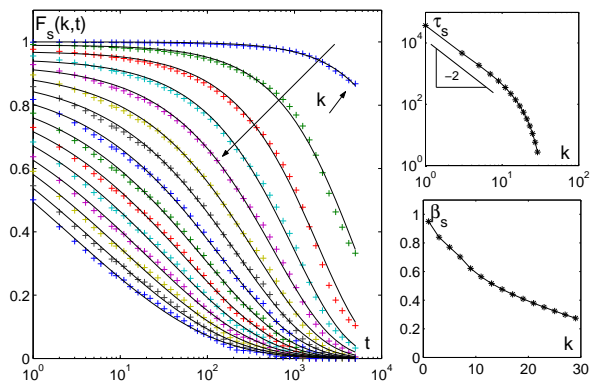}
\caption{On the left: $F_s(k,t)$ as a function of time for different odd
  values of the wave-vector $k=1,3,...,29$ from top to bottom (as indicated by
  the arrow and the increasing $k\nearrow $). The black lines are fits of the form $\exp [-(t/\tau(k))^{\beta(k)}]$. On the right: $\tau(k)$ (top) and
$\beta(k)$ (bottom) as a function of $k$.}
\label{fs}
\vspace{-0.5cm}
\end{figure}
\noindent
The function $F_s(k,t)$ is plotted on the left of Fig.~\ref{fs} as a function
of time for different odd values of $k$ ranging from 1 to 29. Contrary to
glass-formers there is no visible plateau in this correlation function
although from trajectories it is possible to identify a clear cage
effect (see fig.2 of~\cite{Marty}). Note that the short-time dynamics is subdiffusive \cite{Marty}
probably because granular media don't experience thermal relaxation. Therefore the separation of timescales and the corresponding plateau in correlation functions is much less pronounced as in colloids \cite{Weeks}.
Analyzing the curves in Fig.~\ref{fs} we find that the decreasing of $F_s(k,t)$ is slower than exponential in time. A good fit is provided by a stretched exponential: $\exp [-(t/\tau(k))^{\beta(k)}]$. We plot on the right of Fig.~\ref{fs} $\tau(k)$ (top) and $\beta(k)$ (bottom) as a function of $k$.
At small $k$ the relaxation time scales as $k^{-2}$ and the exponent $\beta(k)$ is one. As expected, the grains perform a Brownian motion on large length and time scales and therefore $F_s(k,t)\simeq \exp(-Dk^2t)$ for small $k$ and large $t$.
Increasing $k$ the stretched exponent decreases and is of the order of $0.7$ for $k$ of the order of $2\pi$, corresponding to the inter-grain distance, and even lower for higher values of $k$. A very similar behavior has been found for glass-formers~\cite{KA,Ediger}.
Also the decrease of $\tau (k)$ steepens and decreases sharply for large $k$. This is
also related to the short time sub-diffusive dynamics. In this regime the particle displacement distribution has a variance scaling as $t^{1/2}$ (not $t$ like for standard diffusion) and is well fitted by a Gaussian. Assuming this functional form and Fourier transforming to get the intermediate scattering function, one finds that the relaxation time goes as $k^{-4}$, hence a crossover from a $k^{-2}$ behavior at small $k$ to a more rapid decreasing at large $k$ as in Fig.~\ref{fs}.\\
\begin{figure}
\includegraphics[width=0.72\columnwidth]{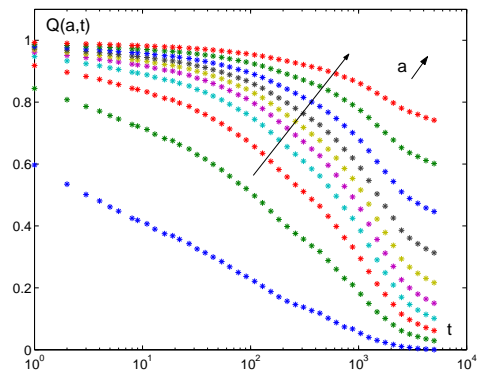}
\caption{$Q_a(t)$ as a function of time for $a=0.05,0.1,...,0.5$.}
\label{q}
\vspace{-0.5cm}
\end{figure}
Dynamical heterogeneity is one of the possible explanation of the non-exponential relaxation of $F_s(k,t)$: the relaxation becomes slower than exponential because there is a strong  spatial distribution of timescales~\cite{Ediger}. This effect is strong for
intermediate and large values of $k$. Instead for smaller $k$s, the heterogeneities are averaged out and a larger value of $\beta(k)$, going to one for $k\rightarrow 0$, is expected. This coincides indeed with the trend found in Fig.~\ref{fs}.
However this is not the only possible scenario \cite{Leticia,Ediger}. In the following we want to go one step further and show direct ``smoking gun'' evidences of dynamical correlations. For this purpose it is of interest to consider the {\it structural relaxation} and not only the single particle one, as given by $F_s(k,t)$. We  
focus on the the density overlap~\cite{FranzParisi} following previous works on glass-forming liquids:
\begin{equation}
Q^{a}=\langle \hat Q^{a}\rangle=\frac{1}{N}\int dr dr' \langle \delta \rho(r,t) w_{a}(r-r') \delta \rho(r',0)
\rangle,
\end{equation}
\noindent
where $\rho(r,t)=\sum_i \delta(r-r_i(t))$ and $\delta \rho(r,t)=\rho(r,t)-\langle \rho \rangle $. The overlap function is a non-normalized Gaussian:
$w_{a}(r)=\exp(-r^2/2a^2)$. The evolution of $Q^a(t)$ is a measure of how long it takes to the systems to decorrelate from its density profile at time $t=0$. Fig.~\ref{q} shows that the behavior of $Q^{a}(t)$ is similar to the one of $F_s(k,t)$, as for glass-formers \cite{FranzParisi,Glotzer}. 
The proper way to unveil spatio-temporal correlations is through the fluctuations of the temporal relaxation~\cite{FranzParisi}. Those are characterized by dynamical susceptibilities:
\begin{eqnarray}\label{xi}
\chi_4^{H}(k,t)&=&N\langle \left(\hat H(k,t)-\langle \hat H(k,t)\rangle\right)^2 \rangle
\end{eqnarray}
\noindent
where $H$ can either be $F_s(k,t)$ or $Q^a(t)$.
They unveil dynamic correlations exactly as fluctuations of the magnetization unveil magnetic correlations close to a ferromagnetic transition, see e.g. \cite{BB,TWBBB}.
One way to understand how such susceptibilities relate to spatial heterogeneities of
the dynamics is to decompose, say, $\hat Q^{a}(t)$ in local
contributions: $N \hat Q^{a}(t)=\rho \int dr \hat q^a(r,t)$ where $\hat
q^a(r,t)=\frac{1}{\rho}\int dr' \delta \rho(r,t) w_{a}(r-r') \delta \rho(r',0)$. Using
this expression one finds $\chi_4^Q(t)=\rho \int dr G_4^{Q}(r,t)$
where $G_4^Q(r,t)=\langle[ \hat q^a(r,t)-\langle \hat q^a(r,t)\rangle][
\hat q^a(0,t)-\langle \hat q^a(0,t)\rangle]\rangle$ is the spatial correlation of the local temporal relaxation : if at point $0$ an event has occurred that leads to a decorrelation of the local density over the time scale $t$, $G_4^Q(r,t)$ is the probability that a similar event has occurred a distance $r$ away, within the same time interval $t$. $\chi_4^Q(t)$, its volume integral, quantifies how much spatially correlated is the dynamics (see \cite{Andersen,Glotzer,BB,TWBBB} for a more detailed discussion).\\
\begin{figure}
\includegraphics[width=0.79\columnwidth]{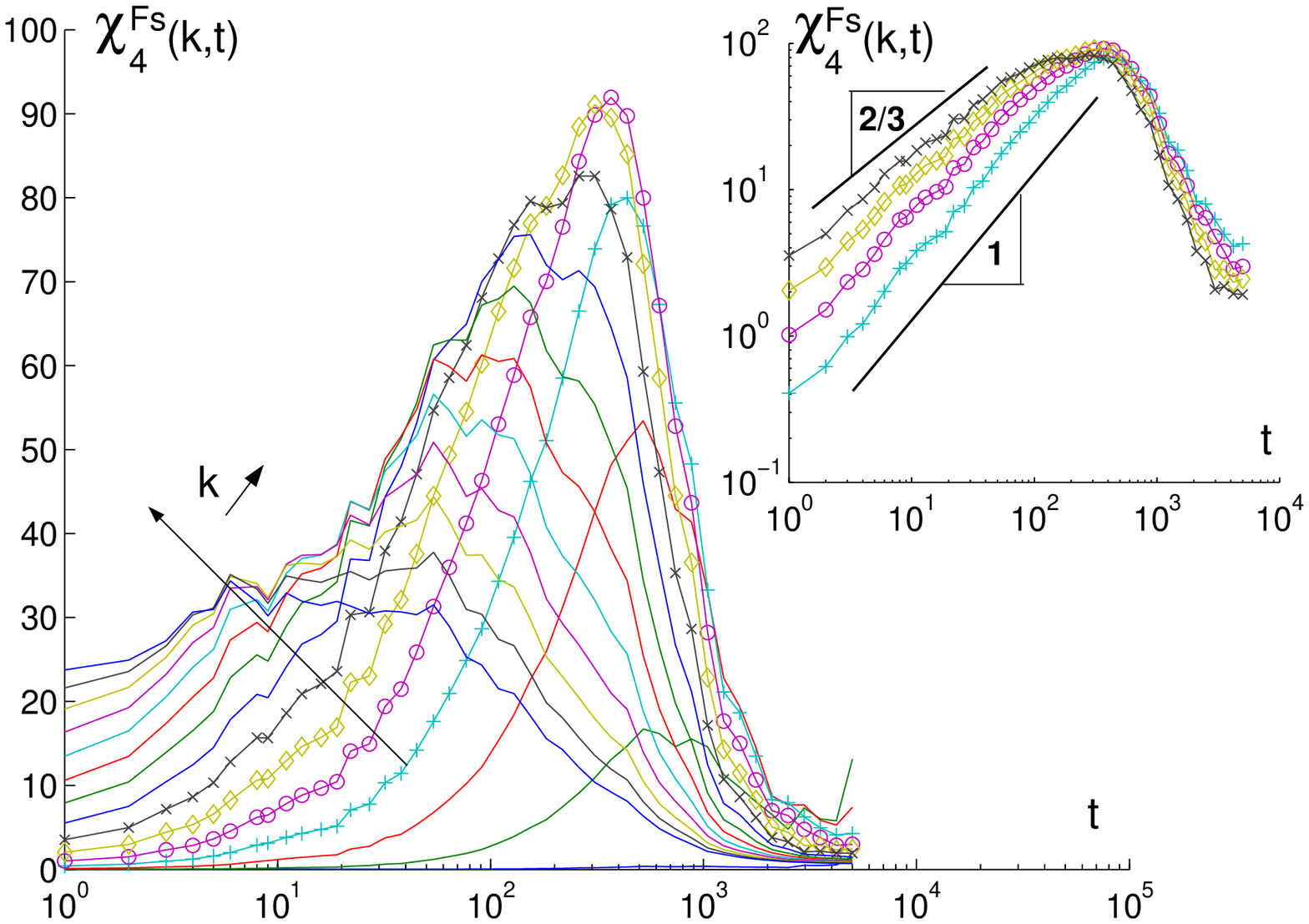}\\
\includegraphics[width=0.79\columnwidth]{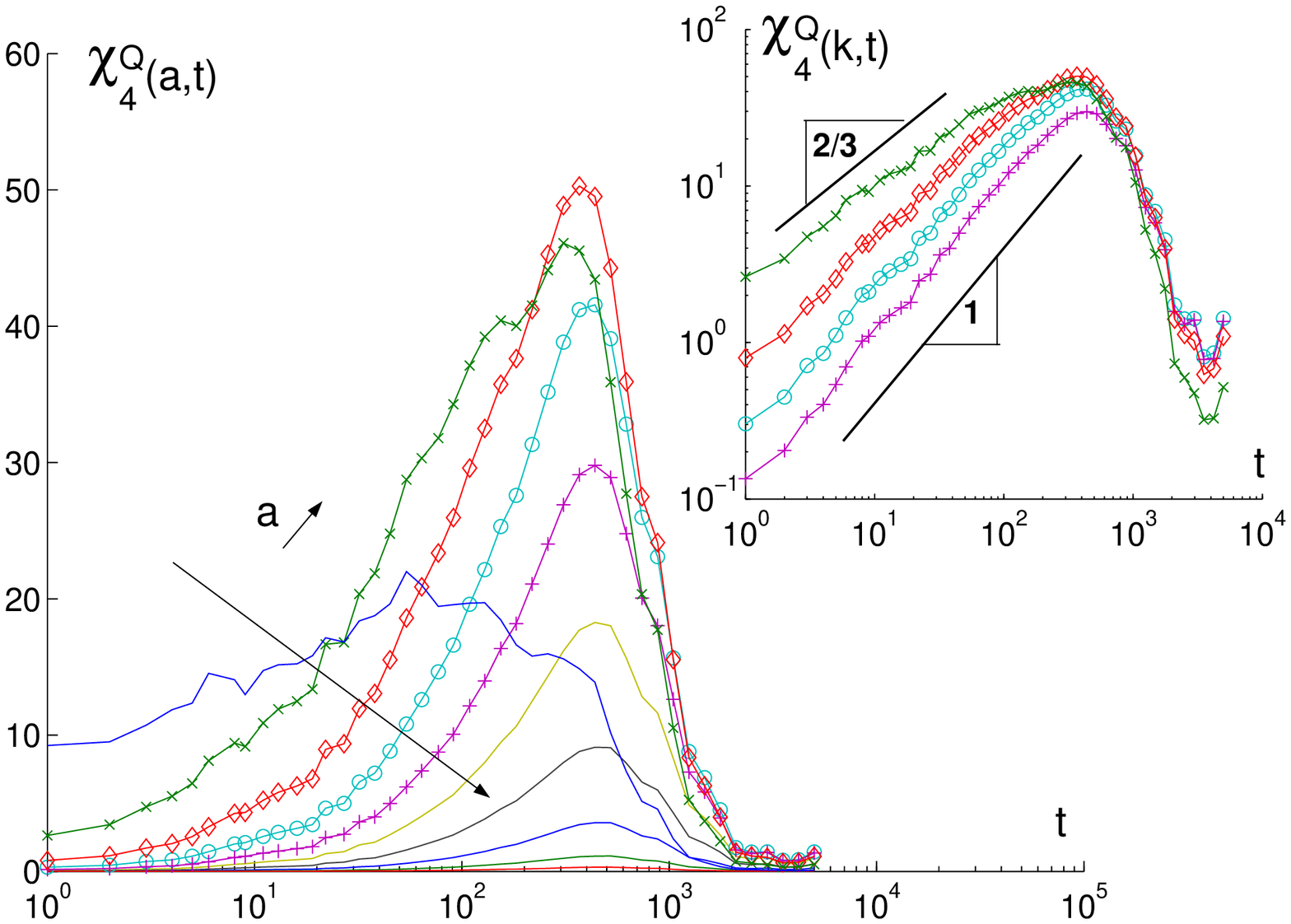}
\caption{Top:$\chi_4^{F_s}(t)$ as a function of time for odd values of $k=1,3,...,29$. 
Inset: Log-Log plot for $k=7,9,11,13$. Bottom:$\chi_4^Q(t)$ as a function of time for values of $a=0.05,0.1,...,0.5$. Inset: Log-Log plot for $a=0.1,0.15,0.2,0.25$.}
\label{chi4fs}
\vspace{-0.5cm}
\end{figure}
\noindent
Top of Fig.~\ref{chi4fs} displays $\chi_4^{F_s}(t)$ for $k=1,3,...,29$. It has the form found for glass-formers \cite{Andersen,Glotzer,Berthier,FranzParisi,TWBBB}: it is of the order of one at small and large time and displays a peak at a time somewhat larger 
than the timescale of the sub-diffusive regime. 
The peak is a clear signature of dynamic heterogeneity and shows that the dynamics is maximally correlated on timescales of the order of the relaxation time. 
A rough estimation of the corresponding dynamical correlation length is obtained identifying the peak of $\chi_4^{F_s}(t)$, of the order of $100$, to a correlated area $\pi \xi_{het}^2$, leading to a length $\xi_{het}\propto 6$ in agreement with a previous estimate~\cite{Marty}. We find very similar results for  $\chi_4^Q(t)$ as shown in the bottom of Fig. \ref{chi4fs} for $a=0.05,0.1,...,0.5$. The largest $\chi_4^{Q}(t)$ is obtained for $a=0.15$, which corresponds to the typical displacement during the sub-diffusive regime. Large and small values of $a$ corresponds to small values of the peak because the dynamics on small lengthscales is certainly not very correlated and on very long lengthscales the heterogeneous character of the dynamics is averaged out. 
As discussed in~\cite{TWBBB}, the power law  growth of $\chi_4^{F_s}(t)$ (resp. $\chi_4^Q(t)$) before the peak with exponents between $1$ and $2/3$ (see insets of Fig.3) suggest that the dynamic correlations cannot be induced by independent defect or free volume diffusion.\\
We now focus on spatio-temporal pattern of mobility. 
\vspace{0.5cm}
\begin{figure}[hb]
\vspace{-0.7cm}
\psfig{file=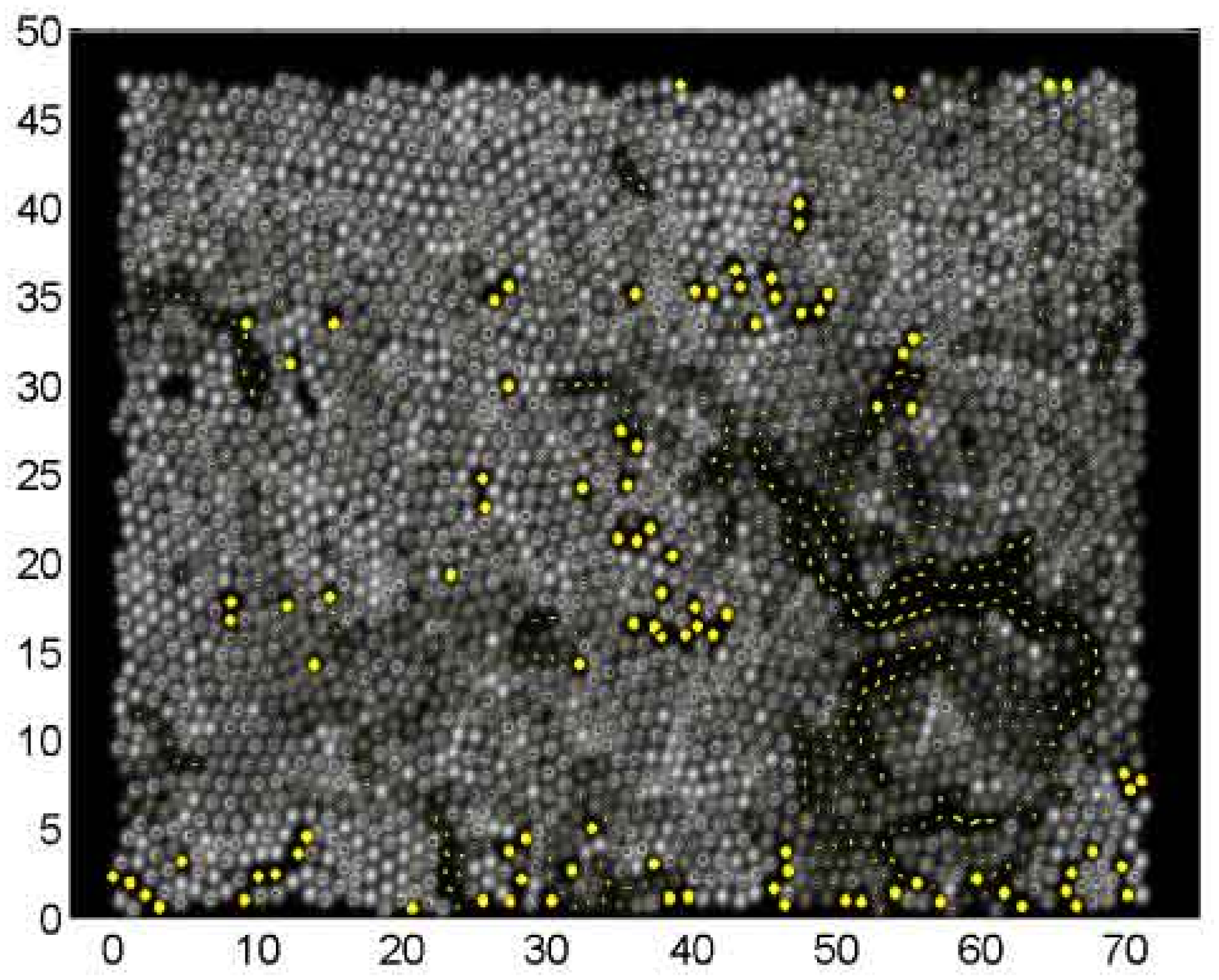,width=6.5cm}\\
\vspace{-0.80cm}
\psfig{file=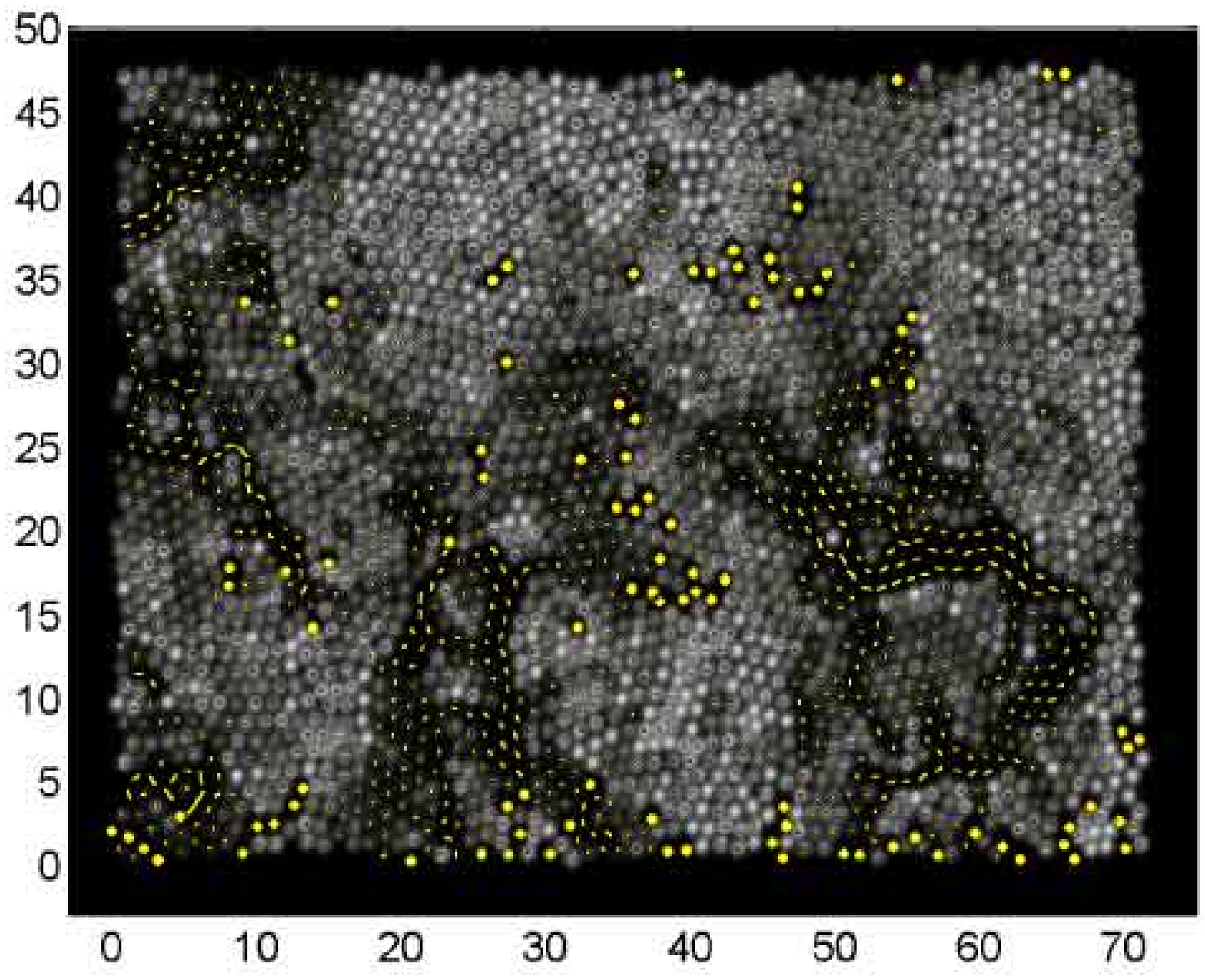,width=6.5cm}\\
\vspace{-0.80cm}
\psfig{file=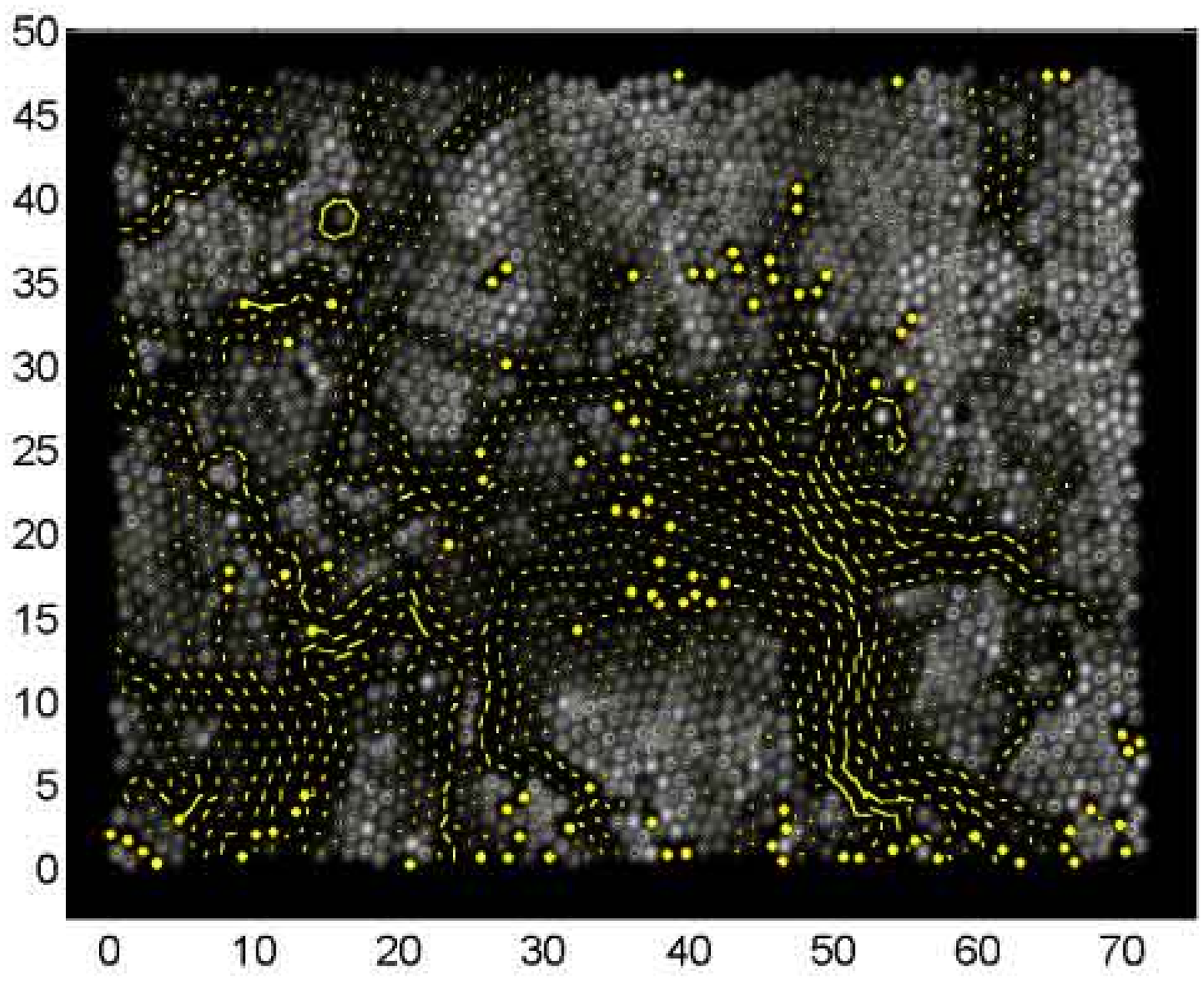,width=6.5cm}\\
\vspace{-0.80cm}
\psfig{file=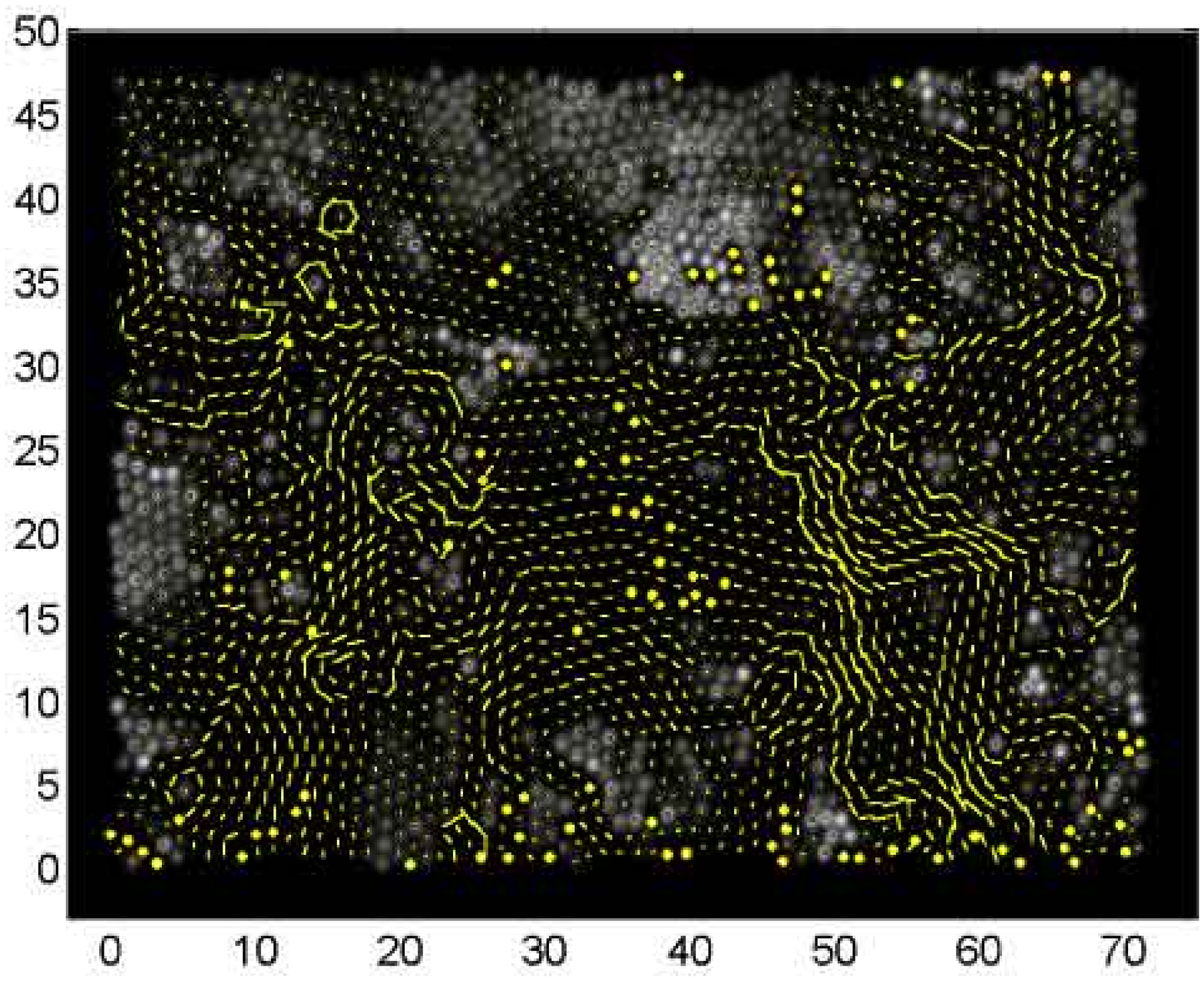,width=6.5cm}\\
\caption{Grey-scale plot of $\hat q^a_s(r,t)$, at $t=42,435,1113,2526$ from top to bottom ($a=0.15$). Black regions correspond to lower values of $\hat q^a_s$. The displacements of the particles during the interval of time $t$ are plotted in yellow. The yellow dots are particles that have been lost during tracking.}\label{mobility}
\end{figure}
\noindent
Fig.~\ref{mobility} presents a grey-scale plot of $\hat q^a_s(r,t)=\sum_i \delta(r-r_i(0))w_a(r_i(t)-r_i(0))$ for $t=42,435,1113,2526$ and $a=0.15$, where  $\delta(r)$ is approximated by a Gaussian of width $0.3$. By definition $\hat q^a_s(r,t)$ measures a coarse grained mobility: if the particle that was close to $r$ at $t=0$ moved away more than $a$ in the time interval $t$ then $\hat q^a_s(r,t)\simeq 0$.  The yellow lines in Fig.\ref{mobility} are the particle displacements in the time interval $t$. At short-times ($t=42$) only few particles have moved and from Fig.\ref{mobility} it appears that they do so in a string-like fashion. On larger times ($t=435,1113$) the relaxed regions are ramified and finally, at very long time ($t=2526$) the majority of the particles has moved substantially but there remain few (rather large) regions not yet relaxed. These findings similar to the ones found in simulation of supercooled liquids~\cite{Andersen,Harrowell,Glotzer,Berthier} suggest that the mobility is organized in clusters, which are the direct visual evidence of the dynamical heterogeneities.
\begin{figure}[h]
\includegraphics[width=0.45\columnwidth]{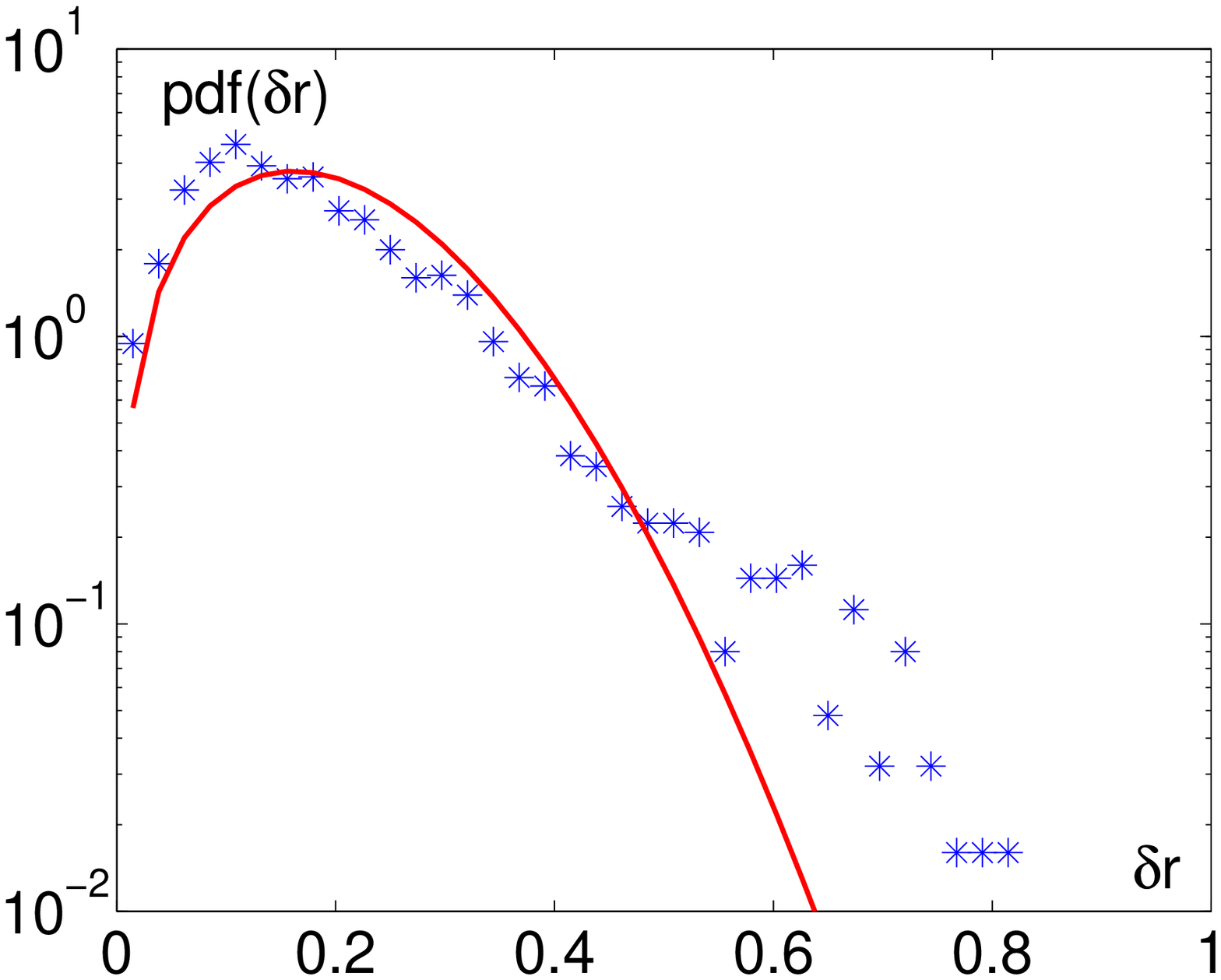}
\hspace{0.2cm}
\includegraphics[width=0.45\columnwidth]{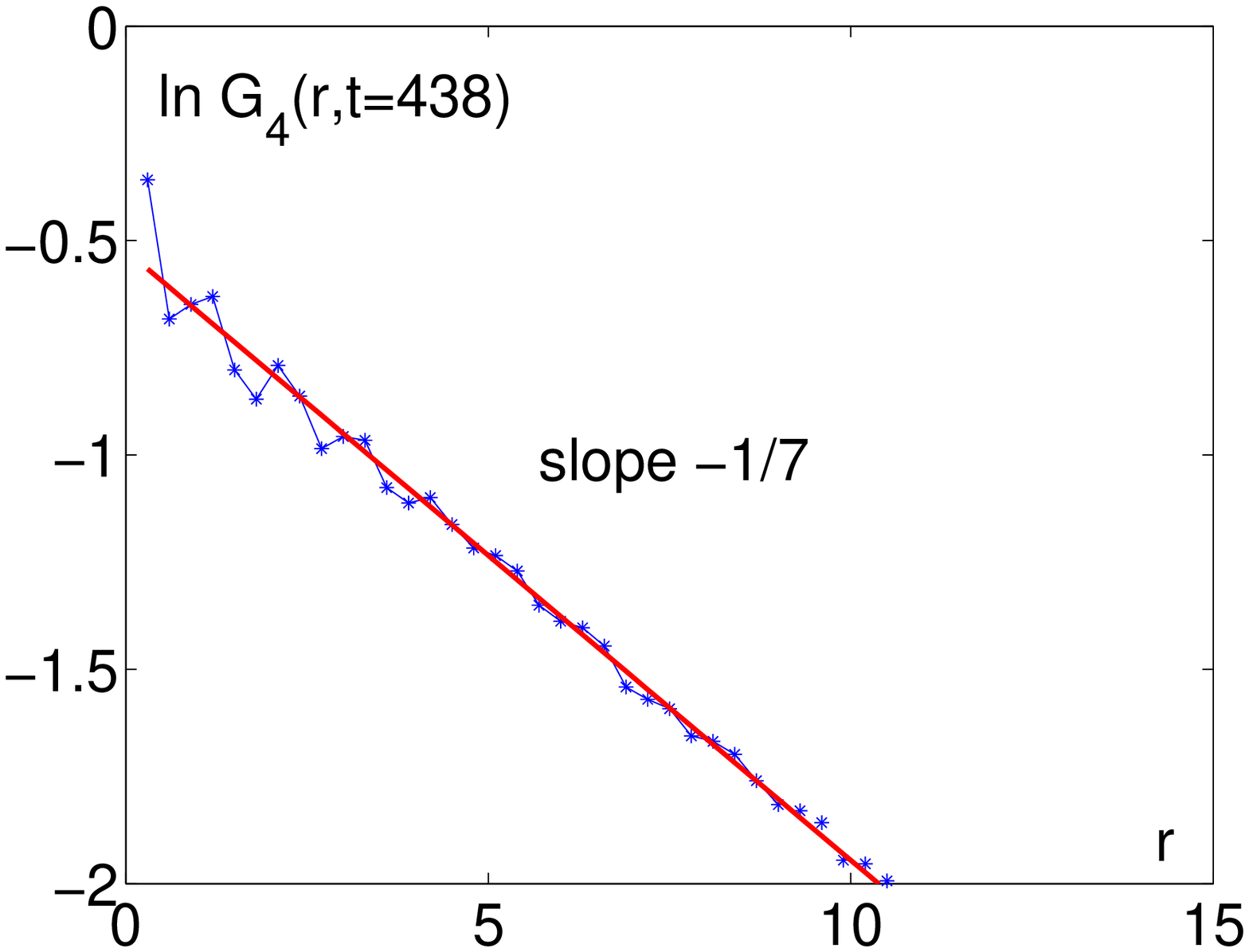}
\caption{Left:self part of the Van Hove correlation function after angular integration at $t=438$; the continuous line is the pdf for a Gaussian distribution. Right: $\ln (G_4(r,438))$ as a function of $r$ (for $a=0.15$); the straight line is a linear fit.}   
\label{G4}
\vspace{-0.5cm}
\end{figure}
\noindent
Left of Fig.~\ref{G4} displays the probability distribution of the grains displacement for $t=438$ (corresponding to the maximum of $\chi_4^Q(t)$) and quantifies the excess of fast and slow grains compared to the Gaussian distribution (in continuous line). 
On the right of Fig.~\ref{G4}, $G_4(r, 438)$, the radial autocorrelation of $q^a_s$, averaged over ten realizations (for $a=0.15$), exhibits an exponential decay over a characteristic dynamical length $\xi=7$, in agreement with the value obtained from the peak of $\chi_4^{F_s}$. For comparison, the dynamic lengthscales reported in experiments close to the glass transition are of the order of $5-10$ molecular diameters \cite{Ediger}.\\
In conclusion our results, to our knowledge the first extensive
experimental study of two and four point {\it spatio-temporal} dynamic correlation functions, furnish a direct experimental evidence that granular materials close to jamming have an heterogeneous and correlated dynamics. 
It would be certainly worth studying the possible relations between the dynamic correlations we have found and the 
diverging length scales that have been proposed to show up at the jamming transition (coming from the jammed phase) \cite{PointJ}. 
Our results reveal a remarkable similarity with glass-forming liquids that reinforces the connection between glasses and jamming
systems \cite{LiuNagel}. They open the way to further analysis, varying a control parameter (as is the temperature for liquids and packing fraction for colloids), or during compaction. That would give other important information on the microscopic dynamics and provide stringent constraints to the theory of glassy and "jammy" materials in general.

\begin{acknowledgments}
We thank all the participants to our "Glassy Working Group" in Saclay and in particular J.-P. Bouchaud 
for enriching and enlightening discussions. We thank C. Gasquet and V. Padilla for technical assistance.
\end{acknowledgments}


\begin{thebibliography}{99}

\bibitem{LiuNagel}
A.J. Liu and S. Nagel, Nature {\bf 396} 21 (1998).

\bibitem{D'Anna}
G. D'Anna and G. Gremaud, Nature {\bf 431} 407 (2001).

\bibitem{Compaction}
J.B. Knight et al., Phys. Rev. E {\bf 51} 3957 (1995).
P. Philippe and D. Bideau, EuroPhys. Lett. {\bf 60} 677 (2002).

\bibitem{aging}
C. Josserand, A. V. Tkachenko, D.M. Mueth, H. Jaeger, Phys. Rev. Lett. 
{\bf 85} 3632 (2000)
A. Kabla and G. Debr{\'e}geas, Phys. Rev. Lett. {\bf 92} 035501.

\bibitem{Pouliquen}
O. Pouliquen, M. Belzons, M. Nicolas, Phys. Rev. Lett. {\bf 91} 014301 (2003).

\bibitem{Marty}
G. Marty and O. Dauchot, Phys. Rev. Lett. {\bf 94} 015701 (2005).

\bibitem{Weeks} E.R. Weeks, J.C. Crocker, A.C. Levitt, A. Schofield, D.A. Weitz, 
Science, {\bf 287}, 627 (2000).

\bibitem{KA}W. Kob and H. C. Andersen
Phys. Rev. E {\bf 52}, 4134-4153 (1995); Phys. Rev. E {\bf 51}, 4626-4641 (1995).

\bibitem{Andersen}
H.C. Andersen, PNAS {\bf 102} 6686 (2005).

\bibitem{Harrowell} M.M. Hurley and
P. Harrowell, Phys. Rev. E {\bf 52}, 1694 (1995).

\bibitem{Glotzer}
N. Lacevic, F.W. Starr, T.B. Schroeder and S.C. Glotzer, J. Chem. Phys. 119 7372 (2003) and refs. therein.

\bibitem{Berthier}
L. Berthier, Phys. Rev. E 69, 020201(R) (2004); S. Whitelam, L. Berthier, and J. P. Garrahan Phys. Rev. Lett. {\bf 92}, 185705 (2004). 

\bibitem{Ediger}
M.A.Ediger Annu. Rev. Phys. Chem. {\bf 51} (2000) 99.

\bibitem{Ball}
P. Ball, Nature {\bf  399}, (1999) 207.

\bibitem{Lefevre} A. Lef{\`e}vre, L. Berthier, R. Stinchcombe, cond-mat/0410741.

\bibitem{Arenzon} J.J. Arenzon, Y. Levin and M. Sellitto,
Physica A {\bf 325} (2003) 371.

\bibitem{FranzParisi}
S. Franz, G. Parisi, J. Phys.:Condens. Matter 12, 6335 (2000).
C. Donati, S. Franz, S.C. Glotzer, G. Parisi,
J. Non-Cryst. Sol., {\bf 307}, 215-224 (2002).

\bibitem{fourpoint} Four point correlation functions, not resolved in space, have been already measured in colloidal systems [A. Duri et al.,  Fluctuation and Noise Letters {\bf 5}, L1 (2005)] and in crystalline materials [Mocuta et al. Science Vol. 308, p.1287 (2005)].

\bibitem{Leticia}
L.F. Cugliandolo, J.L. Iguain, Phys. Rev. Lett. {\bf 85}, 3448 (2000).

\bibitem{BB} G. Biroli and J.-P. Bouchaud, Europhys. Lett. {\bf 67} (2004) 21.

\bibitem{TWBBB} C. Toninelli, M. Wyart, L. Berthier, G. Biroli,
J.-P. Bouchaud, Phys. Rev. E {\bf 71}, 041505 (2005).

\bibitem{HansenMacDonald}
J.P. Hansen and I.R. McDonald, {\it Theory of Simple Liquids}, 2nd edition
, Academic Press, London (1986).


\bibitem{PointJ} L.E. Silbert, A.J. Liu, S.R. Nagel, condmat/0501616 and refs therein;  
Matthieu Wyart, Sidney R. Nagel, T.A. Witten, cond-mat/0409687.
 

\end{thebibliography}
\end{document}